\documentclass[aps,pra,preprint,superscriptaddress]{revtex4-1}
\usepackage{amsfonts}
\usepackage{amsopn}
\usepackage{graphicx}
\begin{document}
%\linenumbers
% Use the \preprint command to place your local institutional report
% number in the upper righthand corner of the title page in preprint mode.
% Multiple \preprint commands are allowed.
% Use the 'preprintnumbers' class option to override journal defaults
% to display numbers if necessary
%\preprint{}

%Title of paper
\title{Quantum union bounds for sequential projective measurements}

% repeat the \author .. \affiliation  etc. as needed
% \email, \thanks, \homepage, \altaffiliation all apply to the current
% author. Explanatory text should go in the []'s, actual e-mail
% address or url should go in the {}'s for \email and \homepage.
% Please use the appropriate macro foreach each type of information

% \affiliation command applies to all authors since the last
% \affiliation command. The \affiliation command should follow the
% other information
% \affiliation can be followed by \email, \homepage, \thanks as well.
\author{Jingliang Gao}
\email[]{gaojingliang@stu.xidian.edu.cn}
%\homepage[]{Your web page}
%\thanks{}
%\altaffiliation{}
\affiliation{State Key Laboratory of Integrated Services Networks, Xidian University, Xi'an 710071, China}
%Collaboration name if desired (requires use of superscriptaddress
%option in \documentclass). \noaffiliation is required (may also be
%used with the \author command).
%\collaboration can be followed by \email, \homepage, \thanks as well.
%\collaboration{}
%\noaffiliation

\date{\today}

\begin{abstract}
We present two new quantum union bounds for sequential projective measurements. These bounds estimate the disturbance accumulation and probability of outcomes when the measurements
are performed sequentially. These results are based on a trigonometric representation of quantum states
and should have wide application in quantum information theory for
information processing tasks such as communication and state discrimination, and perhaps even in the analysis of quantum algorithms.
\end{abstract}

% insert suggested PACS numbers in braces on next line
% insert suggested keywords - APS authors don't need to do this
%\keywords{}

%\maketitle must follow title, authors, abstract, \pacs, and \keywords
\maketitle

% body of paper here - Use proper section commands
% References should be done using the \cite, \ref, and \label commands
% Put \label in argument of \section for cross-referencing
%\section{\label{}}
%\subsection{}
%\subsubsection{}
In order to acquire information from a quantum system, we must perform a quantum measurement on it.
According to quantum theory,
a von Neumann measurement yields an eigenvalue of the measured observable with a probability
given by the Born rule, and simultaneously this measurement could
disturb the measured system. However, by coupling the measuring device to the system weakly,
it is possible to read out certain information while limiting the disturbance to the system~\cite{Jozsa2007}.

In some cases, one can perform a sequence of measurements
in order to acquire the desired information, but
the situation becomes more complex as the number of measurements
increases. Although a single measurement does not necessarily disturb the system in
some cases, the disturbance could potentially accumulate gradually when
the measurements are performed in a sequential fashion.
So some natural questions for sequential measurements are as follows:
{\it Can we bound the accumulated disturbance in a meaningful way
or, related to this, understand how many measurements can be performed
until the final state is no longer close to the initial state?}
Moreover, performing a larger number of measurements
results in a variety of possible sequences. {\it Then how can we estimate
the probability of occurrence of the resulting sequences?}

Having sharp answers to these questions would be very helpful in analyzing
many situations, such as quantum property testing \cite{Ashley2013}, quantum sequential decoding \cite{VG2011,VG2012,Sen2011,Wilde2013},{{ sequential state discrimination\cite{Bergou2013,Pang2013}}}, quantum tomography \cite{Bennett2006},
or any other task which requires a large number of measurements. In former work,
Aaronson presented a union bound for general measurements \cite{Scott2006}. Thereafter,
Sen proposed a significantly improved bound for projective measurements \cite{Sen2011},
his bound now being known as the
``non-commutative union bound.''
Wilde then generalized Sen's bound to apply to general measurements
and analyzed classical communication over a single instance of a quantum channel
with this approach \cite{Wilde2013}.

In this paper, we present some useful bounds for sequential
projective measurements which can be used to estimate the disturbance and
the probability of occurrence separately.
Our results given here strengthen previously known results from \cite{Sen2011} and
\cite{Wilde2013}, and we establish them by employing a trigonometric representation of quantum states. {{As an example of the application, we provide a general formulas for the  sequential decoding strategy\cite{VG2011, VG2012, Sen2011}.}}
%In quantum information theory, a good example is the "sequential decoding strategy" which ueses sequential measurements to decode quantum messages. It is first proposed by VG, and was further developed by Sen and Wilde. VG proved that Holevo bound can be achieved by sequential measurements in \cite{VG2011}\cite{VG2012}. Sen presents a probability theory for sequential projective measurement\cite{Sen2011}. Wilde generalizes Sen's theory to sequential POVM measurement\cite{Wilde2013} and apply it in the quantum polar coding[].

We begin by clarifying what we mean by a sequential measurement. Suppose that the initial state of a quantum
system is given by the density operator $\rho$. Now we perform a sequence of measurements on the system. Specifically, we first perform a two-outcome measurement $\mathcal{M}_{1}$ on $\rho$ and
obtain a post-measurement state $\rho_{1}$. Then we perform another two-outcome
measurement $\mathcal{M}_{2}$ on $\rho_{1}$ and obtain the post-measurement state $\rho_{2}$.
Next, perform a third two-outcome measurement $\mathcal{M}_{3}$ on $\rho_{2}$ and
obtain $\rho_{3}$. And so it carries on, with each measurement being
performed on the state resulting from the previous measurement. After $N$ measurements, we obtain
the state $\rho_{N}$. It should be emphasized that the final state $\rho_{N}$ can take many forms because each step has several possible results. Without loss of generality, we suppose that
each measurement is given by $\mathcal{M}_{i}=\{P_{i}, I-P_{i}\}$ for $ i=1,\ldots N$, where $P_{i}$ are
projectors. (The generality of this approach follows from \cite[Lemma~3.1]{Wilde2013}.)
Now, suppose we are only interested in the case in which each measurement gives the outcome corresponding to $P_{i}$ rather than $I-P_{i}$. In other words, the desired post-measurement state sequence is as follows:
\begin{eqnarray}
&&\rho_{1}=\frac{P_{1}\rho P_{1}}{\operatorname{tr}(P_{1}\rho)}  \nonumber \\
&&\rho_{2}=\frac{P_{2}P_{1}\rho P_{1}P_{2}}{\operatorname{tr}(P_{2}P_{1}\rho P_{1}P_{2})}  \nonumber \\
&&\vdots  \nonumber \\
&&\rho_{N}=\frac{P_{N}...P_{2}P_{1}\rho P_{1}P_{2}...P_{N}}{\operatorname{tr}(P_{N}...P_{2}P_{1}\rho P_{1}P_{2}...P_{N})}\nonumber
\label{eq:1}
\end{eqnarray}

We can now present our main result. The disturbance and the probability of $\rho_{N}$ can
be estimated as stated in the following theorem:\\

\noindent {{\textbf{Theorem 1} \emph{Given a density operator $\rho$ and projectors $P_{1},P_{2},...P_{N}$ such that}
\begin{eqnarray}
\operatorname{tr}(P_{i}\rho)=1-\varepsilon_{i}  \ \ \ i=1,2,...N   \nonumber
\label{eq:2}
\end{eqnarray}
\emph{then we have the following bounds:}\\
(\textbf{1-a}) \emph{ The trace distance between $\rho$ and $\rho_{N}$ obeys}
\begin{eqnarray}
D\big(\rho,\rho_{N}\big)\leq 2\sqrt{\sum\varepsilon_{i}}     \nonumber
\label{eq:3}
\end{eqnarray}
\emph{where $D\big(\rho,\rho_{N}\big)=\operatorname{tr}\sqrt{(\rho-\rho_{N})(\rho-\rho_{N})^{\dagger}}$.}\\
}}

\noindent (\textbf{1-b}) \emph{The probability of the occurrence of $\rho_{N}$ obeys}
\begin{eqnarray}
\operatorname{tr}(P_{N}\cdots P_{2}P_{1}\rho P_{1}P_{2}\cdots P_{N})\geq 1-4\cdot\!\sum\varepsilon_{i}  \nonumber
\label{eq:4}
\end{eqnarray}
{{\emph{The equality holds if and only if all $\varepsilon_{i}$'s are equal to $0$.}}}
\\

%\noindent {\color{blue}{(\textbf{1-c})\emph{For any operator $T$, let $\overline{T}=I-T$}, then
%\begin{eqnarray}
%\overline{P_{1}\cdots P_{N}\cdots P_{1}}\leq 4\sum_{i=1}^{N}\overline{P}_{i}
%\end{eqnarray}

%\emph{Remark1}: Observe that the bound in (4) still holds even without the condition $\sum\varepsilon_{i}\leq\frac{1}{2}$. This is simply because the right side would be negative if $\sum\varepsilon_{i}>\frac{1}{2}$. Furthermore,
%observe that the bounds are really only useful in the case in which each outcome $P_i$ occurs with
%a high probability, as is the case in certain applications in quantum information theory \cite{Sen2011,Wilde2013}.\\
%{\color{blue}{\emph{Remark}: There are some similar results that appeared in former works. First, Wilde proposed a method\cite{Wilde2013} to guarantee that the post-measurement state is close to the original one. He shows that one can perform the projectors $P_{1}$ through $P_{m}$ and then perform them again in the opposite order. The distance between the post-measurement state and the original one can be upper bounded by $\sqrt[4]{\sum\varepsilon_{i}}$. The bound(1-a) improves Wilde's result and reveals that the measurements in opposite order are not necessary. Second, Sen provided a bound which shows that the probability of the occurrence of $\rho_{m}$ is bigger than $1-2\sqrt{\sum \varepsilon_{i}}$. The bound(1-b) obviously improves Sen's bound.}}
{{The bound (1-a) reveals how the disturbance increases when the measurements are sequentially performed. In prior work, Wilde proposed a method\cite{Wilde2013} to guarantee that the post-measurement state is close to the original one. He showed that one can perform the projectors $P_{1}$ through $P_{m}$ and then perform them again in the opposite order. The distance between the post-measurement state and the original one can be upper bounded by $\sqrt[4]{\sum\varepsilon_{i}}$. The bound(1-a) improves Wilde's result and reveals that the measurements in opposite order are not necessary.

{{The bound(1-a) implies that the probability of occurrence of possible results may change by as
much as $O(\sqrt{\sum{\varepsilon_{i}}})$. However, the bound (1-b) provides an even better estimation and controls the change to $O(\sum{\varepsilon_{i}})$. }}It can be thought as a non-commutative analogue of union bound from classical probability theory:
\begin{eqnarray}
\operatorname{Pr}\big\{\overline{A_{1}\!\cap\! \cdots\! \cap\! A_{N}}\big\}\!=\!\operatorname{Pr}\big\{ \overline{A}_{1}\!\cup\! \cdots \!\cup\! \overline{A}_{N}\big\}\!\leq\!\sum_{i=1}^{N}\!\operatorname{Pr}\big\{\overline{A}_{i}\big\} \nonumber
\end{eqnarray}
where $A_{1},\ldots,A_{N}$ are events.
If we think of $P_{1}\!\cdots \!P\!_{N}\!\cdots \!P_{1}$ as the intersection of $P_{i}$'s, then the best analogous bound for projector logic would be
\begin{eqnarray}
1-\operatorname{tr}\left(P_{1}\!\cdots \!P_{N}\!\cdots \!P_{1}\rho\right)\leq \sum_{i=1}^{N}\operatorname{tr}\!\big[\!\left(I-P_{i}\right)\rho\big]  \nonumber
\end{eqnarray}
Though, the above bound only holds if the projectors are commuting. For non-commutative case, the bound(1-b) turns to be the next best thing.

The bound(1-b) can be further generalized as:\\
}}

\noindent {{\textbf{Corollary1:} \emph{For projectors $P_{1},P_{2},...P_{N}$, let $\overline{P}_{i}\!=\!I-P_{i}$, then we have}
\begin{eqnarray}
P_{1}\!\cdots\! P_{N}\!\cdots\! P_{1}\geq I-4\sum_{i=1}^{N}\overline{P}_{i}   \nonumber
\end{eqnarray}
\\
\noindent \textbf{Proof:} This corollary is equivalent to: for any vector $|\nu \rangle$, it holds that
\begin{eqnarray}
\langle\nu|P_{1}\!\cdots\! P_{N}\!\cdots\! P_{1}|\nu\rangle\geq \langle\nu|\nu\rangle-4\sum_{i=1}^{N}\langle\nu|\overline{P}_{i}|\nu\rangle  \nonumber
\label{eq:206}
\end{eqnarray}
Let $\rho=\frac{|\nu\rangle\langle\nu|}{\langle\nu|\nu\rangle}$, then $\rho$ is a density operator. Applying the bound(1-b), the above inequality follows.    $\hfill{} \Box$
%The bound(1-c) can be thought as a non-commutative analogue of union bound from classical probability theory:
%\begin{eqnarray}
%Pr\big\{\overline{A_{1}\cap \cdots \cap A_{N}}\big\}\leq \sum_{i=1}^{N}\!Pr\big\{\overline{A}_{i}\big\}
%\end{eqnarray}
%where $A_{1},\ldots,A_{N}$ are events.
%If we think of $P_{1}\!\cdots \!P\!_{N}\!\cdots \!P_{1}$ as the intersection of $P_{i}$'s, then the exactly analogous bound for projector logic would be
%\begin{eqnarray}
%\overline{P_{1}P_{2}\cdots P\!_{N}\cdots P_{2}P_{1}}\leq  \!\sum_{i=1}^{N}\overline{P}_{i}
%\end{eqnarray}
%Though, the above bound only holds if the projectors are commuting. For non-commutative case, the bound(1-c) turns to the next best thing.
}}

{{\noindent\emph{Remark1:} In prior work, Sen\cite{Sen2011} proved that for any positive operator $\rho$ such that $\operatorname{tr}{\rho}\leq 1$, it holds that
\begin{eqnarray}
\operatorname{tr}(P_{N}\cdots P_{2}P_{1}\rho P_{1}P_{2}\cdots P_{N})\geq \operatorname{tr}\rho-2\sqrt{\!\sum\operatorname{tr}\!\left(\overline{P}_{i}\rho\right)} \nonumber
\end{eqnarray}
The Corollary1 shows that above inequality can be enhanced to the following version:
\begin{eqnarray}
\operatorname{tr}(P_{N}\cdots P_{2}P_{1}\rho P_{1}P_{2}\cdots P_{N})\geq \operatorname{tr}\rho-4\cdot{\!\sum\operatorname{tr}\!\left(\overline{P}_{i}\rho\right)} \nonumber
\end{eqnarray}
The new bound improves Sen's result, particularly in the "Zeno" regime where each measurement succeeds with high probability.}}

In the following, we will detail the proof of Theorem 1. It will be first shown that the bounds hold if $\rho$ is a pure state,
and then extended to the mixed state. Our proof is based on the trigonometric representation of quantum states.

%The () can be rewritten as a simple form,
%\begin{eqnarray}
%tr(P_{m}...P_{2}P_{1}\rho P_{1}P_{2}...P_{m})
%\label{eq:1}
%\end{eqnarray}
Suppose that $\rho=|\psi\rangle\langle\psi|$ is a pure state and the final state is $\rho_{N}=|\psi_{N}\rangle\langle\psi_{N}|$, then we have
\begin{eqnarray}
&|\psi_{1}\rangle&=\frac{P_{1}|\psi\rangle}{\sqrt{\langle\psi|P_{1}|\psi\rangle}}\nonumber\\
&|\psi_{2}\rangle&=\frac{P_{2}|\psi_{1}\rangle}{\sqrt{\langle\psi_{1}|P_{2}|\psi_{1}\rangle}}\nonumber\\
&\vdots&\nonumber\\
&|\psi_{N}\rangle&=\frac{P_{N}|\psi_{N-1}\rangle}{\sqrt{\langle\psi_{N-1}|P_{N}|\psi_{N-1}\rangle}}\nonumber
\label{eq:6}
\end{eqnarray}
Consider the $i$'th measurement, $|\psi_{i}\rangle=\frac{P_{i}|\psi_{i-1}\rangle}{\sqrt{\langle\psi_{i-1}|P_{i}|\psi_{i-1}\rangle}}$. Let $|\psi_{i}^{\bot}\rangle=\frac{(I-P_{i})|\psi_{i-1}\rangle}{\sqrt{\langle\psi_{i-1}|I-P|\psi_{i-1}\rangle}}$, we can write $|\psi_{i-1}\rangle$ in terms of $|\psi_{i}\rangle$ and $|\psi_{i}^{\bot}\rangle$ as follows:
\begin{eqnarray}
|\psi_{i-1}\rangle = \cos{\theta_{i}}|\psi_{i}\rangle+\sin{\theta_{i}}|\psi_{i}^{\bot}\rangle
\label{eq:7}
\end{eqnarray}
where $\theta_{i}=\arccos\Big|\langle\psi_{i}|\psi_{i-1}\rangle\Big|$. \ \ $\theta_{i}$ can be regarded as the angle between $|\psi_{i-1}\rangle$ and $|\psi_{i}\rangle$. The advantage of this representation lies that the trace distance and probability can be expressed in a simple form \cite{Nielsen2000, Wilde2011}:
\begin{eqnarray}
D\big(\psi_{i\!-\!1},\psi_{i}\big)&=&2\sin\theta_{i}\\
\operatorname{tr}(P_{i}|\psi_{i-1}\rangle\langle\psi_{i-1}|)&=&|\langle\psi_{i}|\psi_{i-1}\rangle|^{2}=\cos^{2}\theta_{i}
\label{eq:9}
\end{eqnarray}
If we perform the measurement $\{P_{i}, I-P_{i}\}$ on $\rho$ directly, then the result state would be $|\psi^{\prime}_{i}\rangle=\frac{P_{i}|\psi\rangle}{\sqrt{\langle\psi|P_{i}|\psi\rangle}}$ or $|\psi^{\prime\bot}_{i}\rangle=\frac{(I-P_{i})|\psi\rangle}{\sqrt{\langle\psi|I-P_{i}|\psi\rangle}}$. Likewise, $|\psi\rangle$ can be written as
\begin{eqnarray}
|\psi\rangle=\cos\alpha_{i}|\psi^{\prime}_{i}\rangle+\sin\alpha_{i}|\psi^{\prime\bot}_{i}\rangle
\label{eq:10}
\end{eqnarray}
where $\alpha_{i}=\arccos{\Big|\langle\psi^{\prime}_{i}|\psi\rangle\Big|}$. $\alpha_{i}$ is the angle between $|\psi\rangle$ and $|\psi^{\prime}_{i}\rangle$, and it holds that:
\begin{eqnarray}
D\big(\psi, \psi_{i}^{\prime} \big)&=&2\sin\!\alpha_{i}\\
\operatorname{tr}(P_{i}|\psi\rangle\langle\psi|)&=&\cos^{2}\!\alpha_{i}= \!1\!-\!\varepsilon_{i}
\label{eq:11}
\end{eqnarray}
Thus, we have
\begin{eqnarray}
\sin^{2}\alpha_{i}\!=\!\varepsilon_{i}
\end{eqnarray}
We can also write $|\psi\rangle$ in terms of $|\psi_{i}\rangle$ and its orthogonal complement $|\psi_{i}^{c}\rangle$,
\begin{eqnarray}
|\psi\rangle=\cos\beta_{i}|\psi_{i}\rangle+\sin\beta_{i}|\psi_{i}^{c}\rangle \nonumber
\label{eq:13}
\end{eqnarray}
where $\beta_{i}=\arccos{\Big|\langle\psi_{i}|\psi\rangle\Big|}$. $\beta_{i}$ is the angle between $|\psi\rangle$ and $|\psi_{i}\rangle$, and it holds that:
\begin{eqnarray}
D\big(\psi,\psi_{i}\big)=2\sin\beta_{i}
\label{eq:14}
\end{eqnarray}
\noindent Likewise, let $\gamma_{i}$ be the angle between $|\psi_{i}\rangle$ and $|\psi_{i}^{\prime}\rangle$, then $\gamma_{i}=\arccos{\Big|\langle\psi_{i}|\psi_{i}^{\prime}\rangle\Big|}$.

\begin{figure}
\includegraphics[scale=0.7]{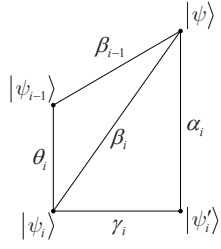}% Here is how to import EPS art
\caption{\label{fig:1} The relationship between the states in the $i$'th measurement}
\end{figure}

For convenience, the states and angles are shown in FIG.\ref{fig:1}. Every vertex in the figure represents a state and the edges indicate the angles.

From the trigonometric representation of the states, we can get two important points. First, from the definition of $\beta_{i}$,
\begin{eqnarray}
\cos{\beta_{i}}&=&|\langle\psi_{i}|\psi\rangle|   \nonumber \\
&=&\left|\cos\alpha_{i}\langle\psi_{i}|\psi^{\prime}_{i}\rangle+\sin\alpha_{i}\langle\psi_{i}|\psi^{\prime\bot}_{i}\rangle\right| \nonumber \\
&=&\cos\alpha_{i}\left|\langle\psi_{i}|\psi^{\prime}_{i}\rangle \right|  \nonumber \\
&=&\cos\alpha_{i}\cos\gamma_{i}
\label{eq:16}
\end{eqnarray}
The equality uses the fact that $P_{i}(I-P_{i})=0$.\\
Second, by (\ref{eq:7})(\ref{eq:10}) we have
\begin{eqnarray}
\cos{\beta_{i-1}}&=&\left|\langle\psi_{i-1}|\psi\rangle\right| \nonumber \\
&=&\left|\cos{\theta_{i}}\cos{\alpha_{i}}\langle\psi_{i}|\psi_{i}^{\prime}\rangle+\sin{\theta_{i}}\sin\alpha_{i}\langle\psi_{i}^{\bot}|\psi_{i}^{\prime\bot}\rangle\right|\nonumber \\
&\leq & \cos\theta_{i}\cos\alpha_{i}\left|\langle\psi_{i}|\psi_{i}^{\prime}\rangle\right|+\sin\theta_{i}\sin\alpha_{i}\nonumber \\
&=&\cos\theta_{i}\cos\alpha_{i}\cos{\gamma_{i}}+\sin\theta_{i}\sin\alpha_{i}
\label{eq:19}
\end{eqnarray}
%So we have
%\begin{eqnarray}
%\cos\!{\theta\!_{i}}\cos\!{\alpha_{i}}\cos{\!\gamma_{i}} \!\geq\! \cos\!{\beta\!_{i-1}}\!-\!\sin\!{\theta\!_{i}}\sin\!{\alpha\!_{i}}
%\label{eq:67}
%\end{eqnarray}
%It is apparent that the right side would be positive when $\alpha_{i}$ and $\beta_{i-1}$ are both small. Precisely, if $\alpha_{i},\beta_{i-1}\leq\frac{\pi}{4}$, then
%\begin{eqnarray}
%\cos\!{\theta\!_{i}}\cos\!{\alpha_{i}}\cos{\!\gamma_{i}} \!\geq\! \cos\!{\beta\!_{i-1}}\!-\!\sin\!{\theta\!_{i}}\sin\!{\alpha_{i}}\!\geq\!0
%\label{eq:20}
%\end{eqnarray}
(\ref{eq:16}) and (\ref{eq:19}) are crucial for our proof of the bounds. We will use them repeatedly in the following.\\

We now prove the following lemma which allows us to lower bound the disturbance in a simple way.\\

{{\noindent \textbf{Lemma 1:} \emph{For the $i$'th measurement, we have}
\begin{eqnarray}
D^{2}\!\big(\psi,\psi_{i}\big)\leq D^{2}\!\big(\psi,\psi_{i-1}\big)+D^{2}\!\big(\psi,\psi^{\prime}_{i}\big)  \nonumber
\label{eq:15}
\end{eqnarray}
\\
\noindent \textbf{Proof:} From the trigonometric representation of the trace distance, this lemma can be equivalently stated as:
\begin{eqnarray}
\sin^{2}{\beta_{i}}\leq\sin^{2}{\beta_{i-1}}+\sin^{2}{\alpha_{i}}  \nonumber
\end{eqnarray}
Furthermore, by(\ref{eq:16}) it is easy to find that
\begin{eqnarray}
\sin^{2}{\beta_{i}}=\cos^{2}{\alpha_{i}}\sin^{2}{\gamma_{i}}+\sin^{2}{\alpha_{i}}  \nonumber
\label{eq:17}
\end{eqnarray}
Therefore, to prove the lemma, we only need to show that: $\sin^{2}\beta_{i-1}\geq \cos^{2}{\alpha_{i}}\sin^{2}{\gamma_{i}}$.\\
%%\begin{eqnarray}
%%\cos^{2}{\alpha_{k}}\sin^{2}{\gamma_{k}}\leq \sin^{2}\beta_{k-1}
%%\label{eq:18}
%%\end{eqnarray}
\noindent Squre (\ref{eq:19}), we have
\begin{eqnarray}
\sin^{2}\!\beta_{i-1}&\geq& 1-\left(\cos\theta\!_{i}\cos\alpha\!_{i}\cos{\gamma\!_{i}}+\sin\theta\!_{i}\sin\alpha\!_{i}\right)^{2} \nonumber\\
&=&\left(\sin\!\theta\!_{i}\cos\!\alpha\!_{i}\cos\!{\gamma\!_{i}}\!+\!\cos\!\theta\!_{i}\sin\!\alpha\!_{i}\!\right)^{2}\!\!+\!\cos^{2}\!{\alpha\!_{i}}\sin^{2}\!{\gamma\!_{i}}\nonumber \\
&\geq& \!\cos^{2}\!{\alpha_{i}}\!\sin^{2}\!{\gamma_{i}} \nonumber
%(\cos\theta_{k}\cos\alpha_{k}\cos\gamma_{k})^{2}\geq (\cos\beta_{k-1}-\sin\theta_{k}\sin\alpha_{k})^2 \nonumber
\label{eq:22}
\end{eqnarray}
This complete the proof. $\hfill{} \Box$ }}\\

Applying Lemma 1, we can obtain that
\begin{eqnarray}
D^{2}\big(\psi,\psi_{N}\big)&\leq&D^{2}\big(\psi,\psi_{N-1}\!\big)+D^{2}\big(\psi,\psi^{\prime}_{N}\big) \nonumber \\
&\leq& D^{2}\big(\psi,\psi_{N-2}\big)+D^{2}\big(\psi,\psi^{\prime}_{N-1}\big)+D^{2}\big(\psi,\psi^{\prime}_{N}\big) \nonumber \\
&\vdots& \nonumber \\
&\leq& \displaystyle{\sum_{i=1}^{N}}D^{2}\big(\psi,\psi^{\prime}_{i}\big)=4\displaystyle{\sum_{i=1}^{N}}\varepsilon_{i}   \nonumber
\label{eq:25}
\end{eqnarray}
Thus, the bound (1-a) is true for pure state.

Now let us consider the case that $\rho$ is mixed state. Suppose that $|\psi\rangle^{RA}$ and $|\psi_{N}\rangle^{RA}$ are purifications of $\rho$ and $\rho_{N}$, where R denotes the reference system. Let $Q_{i}=I^{R}\otimes P_{i}$, then the state $|\psi_{N}\rangle^{RA}$ is generated by performing the projective measurements $\{Q_{i}, I-Q_{i}\}$ sequentially on $|\psi\rangle^{RA}$. Moreover, the probability of each step obeys
\begin{eqnarray}
\operatorname{tr}(Q_{i}|\psi\rangle\langle\psi|^{RA})=\operatorname{tr}(P_{i}\rho)=1-\varepsilon_{i}  \nonumber
%tr\Big(Q_{i}|\psi_{i-1}\rangle\Big\langle\psi_{i-1}|^{RA})=tr\Big(P_{i}|\psi_{i-1}\rangle\langle\psi_{i-1}|\Big)
\label{eq:34}
\end{eqnarray}
Applying the bound for pure state and the monotonicity of trace distance \cite{Nielsen2000,Wilde2011}, we can obtain
\begin{eqnarray}
D\big(\rho, \rho_{N}\big)\leq D\big(\psi^{RA}, \psi_{N}^{RA}\big)\leq 2\sqrt{\sum \varepsilon_{i}}   \nonumber
\label{eq:36}
\end{eqnarray}
This completes the proof of the bound(1-a).\\

The bound (1-b) obviously holds if $\sum\varepsilon_{i}>\frac{1}{2}$ because the right side would be negative. In the following, we will show that it still holds if $\sum\varepsilon_{i}\leq\frac{1}{2}$.

For the pure states, the condition $\sum\varepsilon_{i}\leq\frac{1}{2}$ implies that
\begin{eqnarray}
 0\leq \alpha_{i},\beta_{i}\leq \frac{\pi}{4},\ \   i=1,\ldots,N
\label{eq:205}
\end{eqnarray}
The probability that $|\psi_{N}\rangle$ occurs is
\begin{eqnarray}
&&\ \ \ \operatorname{tr}\big(P_{N}\cdots P_{1}|\psi\rangle\langle\psi| P_{1}\cdots P_{N}\big) \nonumber\\
&&=\operatorname{tr}\big(P_{1}|\psi\rangle\langle\psi|\big)\cdots \operatorname{tr}\big(P_{N}|\psi_{N-1}\rangle\langle\psi_{N-1}|\big)   \nonumber    \\
&&=\cos^{2}\!\theta_{1}\cos^{2}\!\theta_{2}\cdots \cos^{2}\!\theta_{N}
\label{eq:26}
\end{eqnarray}
From(\ref{eq:19}), we can see
\begin{eqnarray}
\cos\beta\!_{N\!-\!1}\!&\leq&\! \cos\theta_{N}\cos\alpha_{N}+\sin\theta_{N}\sin\alpha_{N} \nonumber \\
&=&\cos\left(\theta\!_{N}\!-\!\alpha_{N}\right)  \nonumber
\label{eq:27}
\end{eqnarray}
So it holds that $\theta_{N}\leq \beta_{N\!-\!1}\!+\!\alpha_{N}$. Then we have
\begin{eqnarray}
\cos\!\theta_{1}\!\cdots\! \cos\!\theta_{N}\!\geq\!\cos\!\theta_{1}\!\cdots \!\cos\!\theta\!_{N\!-\!1}\cos\!\left(\beta\!_{N\!-\!1}\!+\!\alpha_{N}\right)
\label{eq:28}
\end{eqnarray}
To continue, we need the following lemma:\\

\noindent\textbf{Lemma 2}: \emph{Define $\{a_{k}\}$ by}
\begin{eqnarray}
a_{k}=\frac{\cos\alpha_{N}\!\cos\beta_{k}\!-\!\!\sqrt{\displaystyle{\sum_{i=k+1}^{N}}\!\!\sin^{2}\!\alpha_{i}}\cdot\sqrt{\sin^{2}\!\beta_{k}\!+\!\displaystyle{\sum^{N-1}_{i=k+1}}\!\!\sin^{2}\!\alpha_{i}}}{1\!+\!\displaystyle{\sum^{N-1}_{i=k+1}}\!\!\sin^{2}\alpha_{i}} \nonumber
\label{eq:29}
\end{eqnarray}
%If the following inequality holds,
%\begin{eqnarray}
%\cos\alpha_{k-1}\leq \cos\theta_{k}\cos\alpha_{k}+\sin\theta_{k}\sin\gamma_{k} \ \ \ \
%\label{eq:1}
%\end{eqnarray}
\emph{Then we have}
\begin{eqnarray}
\cos\theta_{k}\cdot a_{k}\geq a_{k-1} \nonumber
\label{eq:30}
\end{eqnarray}
%\noindent\textbf{Proof:}
Proof: See Appendix A.\\

Note that $\cos\!\left(\beta\!_{N\!-\!1}\!+\!\alpha_{N}\right)=a_{N-1}$. Applying Lemma 2 repeatedly, we can get
\begin{eqnarray}
&&\ \ \ \cos\theta_{1}\cdots \cos\theta\!_{N\!-\!2}\cos\theta\!_{N-1}\cos\left(\beta\!_{N\!-\!1}\!+\!\alpha\!_{N}\right)  \nonumber\\
&&=\cos\theta_{1}\cdots \cos\theta\!_{N\!-\!2}\cos\theta\!_{N\!-\!1}\cdot a\!_{N\!-\!1}  \nonumber\\
&&\geq \cos\theta_{1}\cdots \cos\theta\!_{N-2}\cdot a_{N-2} \nonumber\\
&&\vdots \nonumber \\
&&\geq a_{0}
\label{eq:207}
\end{eqnarray}
Continuing, from the fact that $\beta_{0}=0$, we have
\begin{eqnarray}
a_{0}\!&=&\!\frac{\cos\alpha_{N}\!-\!\sqrt{\displaystyle{\sum_{i=1}^{N}} \sin\!^{2}\!\alpha_{i}\cdot\!\!\displaystyle{\sum_{i=1}^{N-1}}\sin\!^{2}\!\alpha_{i}}}{1+\displaystyle{\sum^{N-1}_{i=1}}\sin^{2}\alpha_{i}}  \nonumber\\
&\geq&\frac{1-\displaystyle{\sum_{i=1}^{N}}\sin^{2}\alpha_{i}}{1+\displaystyle{\sum_{i=1}^{N}}\sin^{2}\alpha_{i}} \\
&=&\frac{1-\sum\varepsilon_{i}}{1+\sum\varepsilon_{i}}\geq 1-2\sum\varepsilon_{i}
\label{eq:31}
\end{eqnarray}
%The second equality follows from the fact that $\beta_{0}=0$. The fourth inequality is shown in Appendix B.\\
The inequality(15) is proved in Appendix B.\\
Combining(\ref{eq:26})(\ref{eq:28})(\ref{eq:207}) and (\ref{eq:31}), we get
\begin{eqnarray}
\operatorname{tr}\Big(\!P_{N}\cdots P_{1}|\psi\rangle\langle\psi| P_{1}\cdots P_{N}\!\Big)\geq 1-4\!\cdot\!\sum\varepsilon_{i} \nonumber
\end{eqnarray}
%Now let us consider the case that $\rho$ is mixed state. Suppose that $|\psi\rangle^{RA}$ and $|\psi_{N}\rangle^{RA}$ are purifications of $\rho$ and $\rho_{N}$, where R denotes the reference system. Let $Q_{i}=I^{R}\otimes P_{i}$, then the state $|\psi_{N}\rangle^{RA}$ is generated by performing the projective measurements $\{Q_{i}, I-Q_{i}\}$ sequentially on $|\psi\rangle^{RA}$. Moreover, the probability of each step obeys
%\begin{eqnarray}
%tr(Q_{i}|\psi\rangle\langle\psi|^{RA})=tr(P_{i}\rho)=1-\varepsilon_{i}
%tr\Big(Q_{i}|\psi_{i-1}\rangle\Big\langle\psi_{i-1}|^{RA})=tr\Big(P_{i}|\psi_{i-1}\rangle\langle\psi_{i-1}|\Big)
%\label{eq:34}
%\end{eqnarray}
%Applying the bound for pure state and the monotonicity of trace distance \cite{Nielsen2000,Wilde2011}, we can obtain
%\begin{eqnarray}
%D\Big(\rho, \rho_{N}\Big)\leq D\Big(|\psi\rangle^{RA}, |\psi_{N}\rangle^{RA}\Big)\leq 2\sqrt{\sum \varepsilon_{i}}
%\label{eq:36}
%\end{eqnarray}
Thus, the bound(1-b) is true for the pure states.

If $\rho$ is a mixed state, then
\begin{eqnarray}
\operatorname{tr}\Big(\!P_{N}\cdots P_{1}\rho P_{1}\!\cdots\! P_{N}\!\Big)\!&=&\!\operatorname{tr}\Big(Q_{N}\!\cdots\! Q_{1}|\psi\rangle\langle\psi|^{RA} Q_{1}\!\cdots\! Q_{N}\!\Big) \nonumber \\
&\geq& 1-4\!\cdot\!\sum\varepsilon_{i} \nonumber
\label{eq:37}
\end{eqnarray}
This completes the proof of bound(1-b).  $\hfill{} \Box$\\

%The bound(1-c) is the generalization of (1-b). To prove it, we need to show that, for any vector $|\nu \rangle$,
%\begin{eqnarray}
%%\langle\nu|\overline{P_{1}\cdots P_{N}\cdots P_{1}}|\nu\rangle\leq 4\sum_{i=1}^{N}\langle\nu|\overline{P}_{i}|\nu\rangle
%\label{eq:206}
%\end{eqnarray}
%Let $\rho=\frac{|\nu\rangle\langle\nu|}{\langle\nu|\nu\rangle}$, then $\rho$ is a density operator. Applying the bound(1-b), (\ref{eq:206}) follows. This
%completes our proof of (1-c).    $\hfill{} \Box$\\

{{Theorem1 reveals how the disturbance accumulates when the measurements are performed sequentially. The generality and simplicity of the bounds imply that they should be nice tools for analyzing many situations. As an example, we will show that how to achieve the Holevo bound via sequential decoding strategy. The sequential decoding scheme was first proposed by Lloyd ,Giovannetti and Maccone(LGM)\cite{VG2011,VG2012}. They showed that it is possible to achieve the Holevo bound by performing sequential measurements. After the work of LGM, Sen presented a simplification of the error analysis by establishing the non-commutative bound\cite{Sen2011}. The new bounds presented in this paper provide a more general formulas for the sequential decoding strategy.

The basic sets of this problem are like this: $\{j\}$ is a set of possible inputs to the quantum channel and $\{\sigma_{j}\}$ are the corresponding outputs. Let $\{p_{j}\}$ be a probability distribution over the indices $\{j\}$ and $\sigma\!\equiv\!\sum p_{j}\sigma_{j}$. Alice wants to send a message chosen from the set $\{1,...,2^{nR}\}$ to Bob by using the quantum channel for $n$ times. The Holvevo bound sets a limit on the rate $R$ that can be achieved when the messages are transferred. We are going to outline a proof that there exists an error-correcting code that accomplishes this task with low probability of error in the limit of large n, and provided $R<S(\sigma)-\sum_{j}p_{j}S(\sigma_{j})$. This proof is based on the random coding and sequential decoding scheme. The transmission of messages can be decomposed into three stages: the encoding, the transmission and the decoding. In the encoding stage, we adopt the standard random coding scheme. To the $i$'th message, Alice associates a codeword $\vec{c}_{i}=c_{1}c_{2}...c_{n}$, where $c_{1},c_{2},...c_{n}$ are chosen from the index set $\{j\}$ according to the distribution $\{p_{j}\}$. She repeats this procedure for $2^{nR}$ times, creating a codebook $\mathcal{C}$ of $2^{nR}$ entries. The corresponding output of the channel is denoted by $\sigma_{\vec{c}_{i}}$. When Bob receives a particular state $\sigma_{\vec{c}_{m}}$ he try to determine what the message was. To do this, he has two tools: the projector $P$ onto the $\delta-$typical subspace of $\sigma^{\otimes n}$ and the projectors $\{P_{\vec{c}_{i}}\}$ onto the $\delta-$typical subspace of corresponding $\sigma_{\vec{c}_{i}}$.  They have the following properties \cite{Wilde2011}: for any $\varepsilon>0$ and sufficiently large n,
\begin{eqnarray}
\operatorname{tr}(P\sigma^{\otimes n})\geq 1-\varepsilon
\label{eq:38}
\end{eqnarray}
\begin{eqnarray}
\operatorname{tr}(P\!_{\vec{c}_{i}}\sigma\!_{\vec{c}_{i}})&\geq& 1-\varepsilon
\label{eq:39}
\end{eqnarray}
\begin{eqnarray}
\operatorname{tr}(P\!_{\vec{c}_{i}})&\leq& 2^{n\left[\sum p_{j}S(\sigma_{j})+\delta\right]}
\label{eq:40}
\end{eqnarray}
\begin{eqnarray}
P\sigma\!^{\otimes n}P &\leq& 2^{-n\left[S(\sigma)-\delta\right]}I
\label{eq:41}
\end{eqnarray}
To decode the message, Bob first performs the measurement $\{P,I-P\}$ to detect whether or not the received state is in the typical subspace of $\sigma^{\otimes n}$. If Yes, he then asks in sequential order, "Is the received codeword $\vec{c}_{i}?$", by performing the measurements $\{P_{\vec{c}_{i}}, I-P_{\vec{c}_{i}}\}$.

The probability of detecting $\vec{c}_{m}$ correctly under this sequential decoding scheme is
\begin{eqnarray}
p_{c}&=&\operatorname{tr}\!\left(\!P_{\vec{c}_{m}}\overline{P}_{\vec{c}_{m\!-\!1}}\!\cdots\! \overline{P}_{\vec{c}_{1}}P\sigma_{\vec{c}_{m}}P\overline{P}_{\vec{c}_{1}}\!\cdots\! \overline{P}_{\vec{c}_{m\!-\!1}}P_{\vec{c}_{m}}\!\right)   \nonumber
%&\geq& tr\left(P\sigma_{\vec{c}_{m}}\right)\!-\!4tr\!\left(\overline{P}_{\vec{c}_{m\!}}P\sigma_{\vec{c}_{m}}P\right)\!-\!4\cdot\!\displaystyle{\sum^{m-1}_{i=1}}tr(P_{\vec{c}_{i}}P\sigma_{\vec{c}_{m}}P) \nonumber
\end{eqnarray}
%where we make the abbreviation $\overline{P}=I-P$. Thus, the error probability is given by
%\begin{eqnarray}
%p_{e}&=&1-tr\!\left(\!P_{\vec{c}_{m}}\overline{P}_{\vec{c}_{m\!-\!1}}\!\cdots\! \overline{P}_{\vec{c}_{1}}P\sigma_{\vec{c}_{m}}P\overline{P}_{\vec{c}_{1}}\!\cdots\! \overline{P}_{\vec{c}_{m\!-\!1}}P_{\vec{c}_{m}}\!\right) \nonumber
%\end{eqnarray}
Consider the expectation of $p_{c}$ over all possible codes $\mathcal{C}$,
\begin{eqnarray}
\mathbb{E}_{\mathcal{C}}\!\{p_{c}\}\!&=&\!\mathbb{E}_{\mathcal{C}}\left\{\operatorname{tr}\!\left(\!P_{\vec{c}_{m}}\overline{P}_{\vec{c}_{m\!-\!1}}\!\cdots\! \overline{P}_{\vec{c}_{1}}P\sigma_{\vec{c}_{m}}P\overline{P}_{\vec{c}_{1}}\!\cdots\! \overline{P}_{\vec{c}_{m\!-\!1}}P_{\vec{c}_{m}}\!\right)\right\}  \nonumber \\
&\geq& \!\mathbb{E}_{\mathcal{C}}\!\Big\{\operatorname{tr}\!\left(P\!\sigma_{\vec{c}_{m}}\!\right)\!-\!4\operatorname{tr}\!\left(\overline{P}_{\vec{c}_{m\!}}\!P\!\sigma_{\vec{c}_{m}}\!P\right)\!-\!4\displaystyle{\sum^{m-1}_{i=1}}\!\operatorname{tr}(P_{\vec{c}_{i}}\!P\sigma_{\vec{c}_{m}}\!P) \!\Big\}  \nonumber
\label{201}
\end{eqnarray}
The inequality follows from Corollary1.\\

For the first term of the right side,
\begin{eqnarray}
\mathbb{E}_{\mathcal{C}}\!\big\{\operatorname{tr}\!\left(P\sigma_{\vec{c}_{m}}\!\right)\!\big\}&=&\operatorname{tr}\!\left(P\mathbb{E}_{\mathcal{C}}\!\{\sigma_{\vec{c}_{m}}\!\}\right) \nonumber \\
&=&\operatorname{tr}(P\sigma^{\otimes n}) \nonumber \\
&\geq&1-\varepsilon \nonumber
\label{eq:202}
\end{eqnarray}
The second equality is due to the fact that $\mathbb{E}_{\mathcal{C}}\!\{\sigma_{\vec{c}_{m}}\!\}=\sigma^{\otimes n}$. The inequality follows from(\ref{eq:38}).

For the second term, we have
\begin{eqnarray}
\mathbb{E}_{\mathcal{C}}\!\Big\{\!\operatorname{tr}\!\left(\overline{P}_{\vec{c}_{m\!}}P\sigma_{\vec{c}_{m}}P\right)\!\Big\}\!
%&&=\!\mathbb{E}_{\mathcal{C}}\!\Big\{tr\Big[\!\sigma\!_{\vec{c}_{m}}\!\!\Big(\overline{P}
%\!_{\vec{c}_{m\!}}\!\!+\!\!\overline{P}\!\!-\!\!\left(PP\!_{\vec{c}_{m\!}}P\!-\!P\!_{\vec{c}_{m\!}}\!\!+\!\!2\overline{P}\right)\!\Big)\!\Big]    \!\Big\} \nonumber \\
%&&\leq\mathbb{E}_{\mathcal{C}}\!\Big\{\! tr\Big[\!\sigma\!_{\vec{c}_{m}}\!\!\Big(\overline{P}
%\!_{\vec{c}_{m\!}}\!\!+\!\!\overline{P}\Big)\!\Big]-\!\!tr\Big(\sigma\!_{\vec{c}_{m}}\!\overline{P}P\!_{\vec{c}_{m\!}}\overline{P}\Big) \!\Big\}  \nonumber \\
%&&\leq\mathbb{E}_{\mathcal{C}}\!\Big\{\! tr\Big(\!\sigma\!_{\vec{c}_{m}}\!\overline{P}
%\!_{\vec{c}_{m\!}}\Big) \!\Big\}+\mathbb{E}_{\mathcal{C}}\!\Big\{\!tr\Big(\sigma\!_{\vec{c}_{m}}\overline{P}\Big) \!\Big\}  \nonumber\\
%&&\leq2\varepsilon
&=&\mathbb{E}_{\mathcal{C}}\!\Big\{\operatorname{tr}\!\left(P\sigma_{\vec{c}_{m}}\right)\!-\!\operatorname{tr}\!\left(P_{\vec{c}_{m\!}}P\sigma_{\vec{c}_{m}}P\right) \Big\} \nonumber \\
&\leq&1\!-\!\mathbb{E}_{\mathcal{C}}\!\Big\{\operatorname{tr}\!\left(P_{\vec{c}_{m\!}}P\sigma_{\vec{c}_{m}}P\right)\Big\}\nonumber \\
&\leq&4\cdot\mathbb{E}_{\mathcal{C}}\Big\{\!\operatorname{tr}\left(\overline{P}\sigma_{\vec{c}_{m}}\right)\!+\!\operatorname{tr}\left(\overline{P}\!_{\vec{c}_{m\!}}\sigma\!_{\vec{c}_{m}}\right)\!\Big\}  \nonumber \\
&\leq&8\varepsilon \nonumber
\label{eq:203}
\end{eqnarray}
The first inequality uses the fact $tr\!\big(P\sigma_{\vec{c}_{m}}\big)\!\!\leq\!\! 1$. The second inequality is due to the bound(1-b). The last inequality follows from(\ref{eq:38}) and (\ref{eq:39}).

For the third term, we have
\begin{eqnarray}
\mathbb{E}_{\mathcal{C}}\!\Big\{\displaystyle{\sum^{m-1}_{i=1}}\!\operatorname{tr}(P_{\vec{c}_{i}}\!P\sigma_{\vec{c}_{m}}\!P)\Big\}&\leq&\mathbb{E}_{\mathcal{C}}\!\Big\{\displaystyle{\sum_{i\neq m}}\!\operatorname{tr}(P_{\vec{c}_{i}}\!P\sigma_{\vec{c}_{m}}\!P)\Big\}\nonumber \\
%&=&\displaystyle{\sum_{i\neq m}}tr\left(\mathbb{E}_{\mathcal{C}}\!\{P_{\vec{c}_{i}}\}P\mathbb{E}_{\mathcal{C}}\{\sigma_{\vec{c}_{m}}\}\!P\right) \nonumber \\
&=&\displaystyle{\sum_{i\neq m}}\operatorname{tr}\left(\mathbb{E}_{\mathcal{C}}\!\{P_{\vec{c}_{i}}\}P\sigma^{\otimes n}P\right) \nonumber \\
%&\leq&\frac{2\!^{-n[S(\sigma)-\delta]}}{1-\varepsilon}\displaystyle{\sum_{i\neq m}}tr\Big(\mathbb{E}_{\mathcal{C}}\{P_{\vec{c}_{i}}\}P\Big) \nonumber \\
%&=&2\!^{-n[S(\sigma)-\delta]}\displaystyle{\sum_{i\neq m}}\mathbb{E}_{\mathcal{C}}\{tr(P_{\vec{c}_{i}}P)\} \nonumber \\
&\leq&2\!^{-n[S(\sigma)-\delta]}\cdot\!\displaystyle{\sum_{i\neq m}}\mathbb{E}_{\mathcal{C}}\{tr(P_{\vec{c}_{i}})\} \nonumber \\
&\leq&2\!^{-n[S(\sigma)\!-\!\delta]}\!\cdot\!\left(\!2^{nR}\!-\!1\!\right)\!\cdot\!2^{n\left[\sum\!\!p_{i}\!S(\!\sigma\!_{i}\!)+\delta\right]} \nonumber \\
&<&2^{n\left[R-(\chi-2\delta)\right]} \nonumber
\label{eq:204}
\end{eqnarray}
where $\chi=S(\sigma)\!-\!\sum\! p_{j}S(\sigma\!_{j})$ is the Holevo quality. The first inequality follows from summing all of the codewords not equal to $\vec{c}_{m}$(this sum can only be larger). The second inequality is due to (\ref{eq:41}). The third inequality follows from (\ref{eq:40}).

Thus, the average probability we get the correct result turns to be
\begin{eqnarray}
\mathbb{E}_{\mathcal{C}}\left\{p_{c}\right\}>1-33\varepsilon-4\!\cdot\!2^{n\left[R-(\chi-2\delta)\right]} \nonumber
\label{eq:51}
\end{eqnarray}
The error probability $p_{e}=1-p_{c}$, so
\begin{eqnarray}
\mathbb{E}_{\mathcal{C}}\left\{p_{e}\right\}< 33\varepsilon+4\!\cdot\!2^{n\left[R-(\chi-2\delta)\right]} \nonumber
\label{eq:52}
\end{eqnarray}
It means that there exists at least one code such that
\begin{eqnarray}
p_{e}< 33\varepsilon+4\!\cdot\!2^{n\left[R-(\chi-2\delta)\right]} \nonumber
\end{eqnarray}
$\varepsilon$ and $\delta$ can be arbitrary small, so for any $R$ such that $R<\chi$, $p_{e}\rightarrow0$ when $n\rightarrow\infty$. This completes our proof.}}\\

{{\noindent\emph{Remark2:} Sen also provided a similar decoding procedure in\cite{Sen2011}. In his proof, the expected error probability  is
\begin{eqnarray}
p_{e}<2\sqrt{4\cdot2^{n\left[R-(\chi-2\delta)\right]}+13\sqrt{\varepsilon}}
\label{eq:105}
\end{eqnarray}
We can see that, the error analysis that we have shown above is significantly better than Sen' result.\\

\noindent\emph{Remark3}:  It would be interesting to compare the Corollary1 with the Hayashi-Nagaoka inequality\cite{Hayashi2003} which plays the key role in "prettry good measurement". In the pretty good measurement, the detecting operator of $\vec{c}_{m}$ is defined by
\begin{eqnarray}
\Lambda_{m}^{p}=\bigg(\!\sum_{i}PP\!_{\vec{c}_{i}}P\!\bigg)^{-\frac{1}{2}}PP\!_{\vec{c}_{m}}P\bigg(\!\sum_{i}PP\!_{\vec{c}_{i}}P\!\bigg)^{-\frac{1}{2}}
\end{eqnarray}
The error probability can be bounded by applying the Hayashi-Nagaoka inequality
\begin{eqnarray}
(S+T)^{-\frac{1}{2}}S(S+T)^{-\frac{1}{2}}\geq I-2(I-S)-4T
\end{eqnarray}
Let $S\!=\!PP\!_{\vec{c}_{m}}\!P,\  T\!=\!\displaystyle{\sum_{i\neq m}}\!PP\!_{\vec{c}_{i}}P$, then
\begin{eqnarray}
\Lambda_{m}^{p}\geq P-2P\overline{P}_{\vec{c}_{m}}\!P-4\displaystyle{\sum_{i\neq m}}\!PP\!_{\vec{c}_{i}}P
\label{eq:301}
\end{eqnarray}
In our sequential decoding scheme, the detecting operator of $\vec{c}_{m}$ is
\begin{eqnarray}
\Lambda_{m}^{s}=P\overline{P}_{\vec{c}_{1}}\!\cdots\! \overline{P}_{\vec{c}_{m\!-\!1}}P_{\vec{c}_{m}}\overline{P}_{\vec{c}_{m\!-\!1}}\!\cdots\! \overline{P}_{\vec{c}_{1}}P
\end{eqnarray}
Applying the Corollary1, then we have
\begin{eqnarray}
\Lambda_{m}^{s}\geq P-4P\overline{P}_{\vec{c}_{m}}\!P-4\displaystyle{\sum_{i=1}^{m-1}}\!PP\!_{\vec{c}_{i}}P
\label{eq:302}
\end{eqnarray}
We can see that, the Corollary1 actually plays a similar role as that the Hayashi-Nagaoka inequality plays in pretty good measurement and they give very similar error analysis.}}\\

\noindent\emph{Conclusion:} With the aid of the trigonometric representation of quantum states, we find two union bounds for estimating the disturbance and probability of the sequential projective measurements. Our result provides a powerful tool for analyzing many situations. As an example, we provide a new proof of achieving the Holveo Bound via sequential measurements.

{{It is not clear to us whether the bounds still hold for sequential POVMs, or stronger, for sequential general measurements. It would be an interesting open problem for further study. What we have known so far is that, the bound(1-b) holds when we perform the same POVM repeatedly, i.e., if $tr(E\rho)=1-\varepsilon$, then $tr(E^{m}\rho)>1-m\varepsilon$. This is a simple consequence of the quantum Jensen inequality.}}

\begin{acknowledgments}
I thank Mark. M. Wilde and N. Cai for useful discussions. This work is supported by the National Natural Science Foundation of China Grant No.61271174, No.61372076, No.61301178£¬No.61301172, No. 61502376 and the 111 Program of China under Grant No. B08038.
\end{acknowledgments}

\appendix

{{\section{}
\noindent {{In this appendix, the proof of Lemma2 is specified.}} From(\ref{eq:16})(\ref{eq:19}) and (\ref{eq:205}), we have
\begin{eqnarray}
\cos\beta_{k}\cos\theta_{k}\geq \cos\beta_{k-1}-\sin\theta_{k}\sin\alpha_{k}\geq0  \nonumber
\label{eq:58}
\end{eqnarray}
Let $x=\sin\theta_{k}$, then from the definition of $a_{k}$, we have
\begin{widetext}
\begin{eqnarray}
&&\cos\theta_{k}\cdot a_{k}=\frac{\cos\alpha_{N}\cos\beta_{k}\cos\theta_{k}-\sqrt{\displaystyle{\sum_{i=k+1}^{N}}\sin^{2}\alpha_{i}}\cdot\sqrt{\cos^{2}\theta_{k}-\cos^{2}\beta_{k}\cos^{2}\theta_{k}+\cos^{2}\theta_{k}\displaystyle{\sum^{N-1}_{i=k+1}}\sin^{2}\alpha_{i}}}{1+\displaystyle{\sum^{N-1}_{i=k+1}}\sin^{2}\alpha_{i}} \nonumber \\
&&\geq \frac{\cos\alpha_{N}\left(\cos\beta_{k-1}-\sin\theta_{k}\sin\alpha_{k}\right)-\sqrt{\displaystyle{\sum_{i=k+1}^{N}}\sin^{2}\alpha_{i}}\cdot\sqrt{\cos^{2}\theta_{k}-\left(\cos\beta_{k-1}-\sin\theta_{k}\sin\alpha_{k}\right)^{2}+\cos^{2}\theta_{k}\displaystyle{\sum^{N-1}_{i=k+1}}\sin^{2}\alpha_{i}}}{1+\displaystyle{\sum^{N-1}_{i=k+1}}\sin^{2}\alpha_{i}} \nonumber  \\
&&= \frac{\cos\!\alpha_{N}\!\cos\!\beta_{k\!-\!1}\!-\!x\cos\!\alpha_{N}\sin\!\alpha_{k}\!-\!\!\sqrt{\displaystyle{\sum_{i=k+1}^{N}}\!\sin^{2}\!\!\alpha_{i}}\!\cdot\! \sqrt{\!-\!\!\left(\!1\!+\!\!\displaystyle{\sum_{i=k}^{N-1}}\!\!\sin\!^{2}\!\alpha_{i}\!\right)\!x^{2}\!+\!2x\cos\beta\!_{k-1}\!\sin\!\alpha\!_{k}\!+\!\!\!\displaystyle{\sum_{i=k+1}^{N-1}}\!\!\!\sin\!^{2}\!\alpha_{i}\!+\!\sin\!^{2}\!\beta_{k-1} }}{1+\displaystyle{\sum^{N-1}_{i=k+1}}\sin^{2}\alpha_{i}} \nonumber
\label{eq:59}
\end{eqnarray}
\end{widetext}
Denote the right side by $g(x)$.
%\begin{widetext}
%\begin{eqnarray}
%g(x)=\frac{\cos\alpha_{m}\cos\beta_{k-1}-x\cos\alpha_{m}\sin\alpha_{k}-\sqrt{\displaystyle{\sum_{i=k+1}^{m}}\sin^{2}\alpha_{i}}\cdot \sqrt{-\left(1-\displaystyle{\sum_{i=k}^{N-1}}\sin^{2}\alpha_{i}\right)x^{2}+2x\cos\beta_{k-1}\sin\alpha_{k}+\displaystyle{\sum_{i=k+1}^{N-1}}\sin^{2}\alpha_{i}+\sin^{2}\beta_{k-1} }}{1+\displaystyle{\sum^{m-1}_{i=k+1}}\sin^{2}\alpha_{i}}         \nonumber
%\label{eq:1}
%\end{eqnarray}
%\end{widetext}
From $g'(x)=0$, we can obtain the minimum value of $g(x)$. It can be verified that $g_{min}(x)=a_{k-1}$ iff
\begin{widetext}
\begin{eqnarray}
x=\frac{\cos\beta_{k-1}\sin\alpha_{k}\left(\displaystyle{\sum_{i=k}^{N}}\sin^{2}\!\alpha_{i}\right)+\sin\alpha_{k}\cos\alpha_{N}\sqrt{\displaystyle{\sum_{i=k}^{N}}\sin^{2}\!\alpha_{i}}\cdot\sqrt{\left(\sin^{2}\beta_{k-1}+\displaystyle{\sum_{i=k}^{N-1}}\sin^{2}\!\alpha_{i}\right)}}{\left(1+\displaystyle{\sum_{i=k}^{N-1}}\sin^{2}\!\alpha_{i}\right)\left(\displaystyle{\sum_{i=k}^{N}}\sin^{2}\!\alpha_{i}\right)}
\nonumber
\label{eq:100}
\end{eqnarray}
\end{widetext}
}}
{{\section{}
\noindent {{To prove the inequality(15)}}, we first define $W$ by
\begin{eqnarray}
W\!=\!\cos\alpha_{N}\!-\!\sqrt{\displaystyle{\sum_{i=1}^{N}}\sin^{2}\!\alpha_{i}\displaystyle{\sum_{i=1}^{N-1}}\!\sin^{2}\!\alpha_{i}\!}-\! \left(\!1\!-\!\displaystyle{\sum_{i=1}^{N}}\sin^{2}\!\alpha_{i}\!\right)  \nonumber
\label{eq:62}
\end{eqnarray}
Clearly, if $W\geq0$, then the inequality holds.
It can be verified that
\begin{eqnarray}
W\!=\!\frac{\sin^{2}\alpha_{N}}{1\!+\!\sqrt{1\!-\!{\frac{\sin^{2}\alpha_{N}}{\sum\sin^{2}\alpha_{i}}}}}-\frac{\sin^{2}\alpha_{N}}{1\!+\!\sqrt{1\!-\!\sin^{2}\alpha_{N}}} \nonumber
\label{eq:63}
\end{eqnarray}
Since $\sum\!\sin^{2}\!\!\alpha_{i}\!\!=\!\!\sum\varepsilon_{i}\!\leq\!\frac{1}{2}$, we have $W\geq0$. $\hfill{} \Box$}}

\bibliography{ref1}

\end{document}